\documentclass{aa}
\usepackage{graphicx}
\graphicspath{{figures/}}
\usepackage{newtxtext, newtxmath}
\usepackage[colorlinks=true, allcolors=blue]{hyperref}
\usepackage{orcidlink}
\usepackage[locale=US]{siunitx}
\DeclareSymbolFont{upgreek}{U}{eur}{m}{n}
\DeclareMathSymbol{\umu}{0}{upgreek}{"16}
\DeclareSIPrefix{\micro}{\text{\ensuremath{\umu}}}{-6}
\let\oldbibliography\thebibliography
\renewcommand{\thebibliography}[1]{%
    \oldbibliography{#1}%
    \setlength{\itemsep}{0pt}%
}
\begin{document}

\title{Feasibility study on retrieving exoplanetary cloud cover distributions using polarimetry}

\titlerunning{Retrieving cloud cover distributions using polarimetry}


\author{
    S. Winning\,\orcidlink{0009-0006-9992-7405}\and
    M. Lietzow-Sinjen\,\orcidlink{0000-0001-9511-3371}\and
    S. Wolf\,\orcidlink{0000-0001-7841-3452}
    }

\authorrunning{S. Winning\space\etalname}

\institute{
    Institute of Theoretical Physics and Astrophysics,
    Kiel University, Leibnizstr. 15, 24118 Kiel, Germany\\
    \email{simon.winning@stu.uni-kiel.de}
    }

\date{Received; accepted}

\abstract
{As a new growing field, exocartography aims to map the surface features of exoplanets that are beyond the resolution of traditional observing techniques.
While photometric approaches have been discussed extensively, polarimetry has received less attention despite its promising prospects.}
{We demonstrate that the limb polarization of an exoplanetary atmosphere offers valuable insights into its cloud cover distribution.
Specifically, we determine an upper limit for the polarimetric precision, which is required to extract information about the latitudinal cloud cover of temperate Jovian planets for scenarios of observations with and without host stars.}
{To compute the scattered stellar radiation of an exoplanetary atmosphere and to study the polarization at various planetary phase angles, we used the three-dimensional Monte Carlo radiative transfer code POLARIS.}
{When the planetary signal can be measured separately from the stellar radiation, information about the latitudinal cloud cover for polar cap models is accessible at polarimetric sensitivities of \SI{0.1}{\percent}.
In contrast, a precision of about \SI{e-3}{ppm} is required when the stellar flux is included to gain this information.}
{}


\maketitle

\nolinenumbers


\section{Introduction}
\label{sec:introduction}

In recent decades, the pursuit of exocartography has gained considerable momentum.
Exocartography aims to unveil the surface features on exoplanets that are otherwise beyond the resolving capabilities of conventional observation techniques.
While photometric strategies have been discussed in detail, polarimetric approaches are still less well explored.
In this study, we highlight a specific aspect of polarimetric mapping techniques by discussing the feasibility and difficulties of retrieving information about the latitudinal cloud cover from the polarization of the phase-angle-dependent scattered stellar radiation of extrasolar planets.

Summarized by \citet{cowan-fujii-2018}, numerous strategies have found their way into the widely accepted toolkit of exocartography in recent decades.
Different photometric considerations show that the radiation emitted by planets is sensitive to their obliquity and surface conditions \citep{gaidos-williams-2004, cowan-etal-2012, cowan-etal-2013, gomez-leal-etal-2012}.
In particular, by investigating the thermal emission of $\upsilon$~Andromedae~b obtained with the Spitzer Space Telescope, \citet{harrington-etal-2006} were able to determine a temperature difference between the day- and nightside of the planet from the phase-dependent flux.
Phase variations due to the contrast between the day- and nightside of an exoplanet were also found, for instance, for HD~179949~b \citep{cowan-etal-2007} and CoRoT-1~b \citep{snellen-etal-2009}.
Furthermore, the spatial variation in the thermal emission can also be retrieved from secondary eclipse light curves of transiting exoplanets \citep{williams-etal-2006, rauscher-etal-2007}.
In particular, observations of the transits and eclipses of HD~189733~b \citep{knutson-etal-2007, knutson-etal-2009, agol-etal-2010} and Kepler-7~b by \citep{demory-etal-2013} revealed an offset of the temperature hotspot on the dayside of the planet.
Subsequently, a map of the brightness distribution of the dayside of the planet was created \citep{majeau-etal-2012, majeau-etal-2012-erat, de-wit-etal-2012}.

In addition to the emitted radiation, reflected light also contains information about the orbital parameters, the rotation rate, and the obliquity of the planet \citep{palle-etal-2008, kawahara-2016, schwartz-etal-2016, nakagawa-etal-2020}.
\citet{cowan-agol-2008} described an inversion technique for creating brightness maps for tidally locked exoplanets from the reflected flux.
In addition, \citet{cowan-etal-2009} investigated disk-integrated and time-averaged light curves of Earth to test whether oceans and continents can be detected, and \citet{oakley-cash-2009} used the reflected light curve to determine a change in obliquity or seasonal terrain.
Based on the method by \citet{fujii-etal-2010} for reconstructing the fractional areas of different surface types such as water oceans, soil, snow, and vegetation, \citet{kawahara-fujii-2010} developed a method for retrieving land distribution with longitudinal and latitudinal resolutions.
\citet{kawahara-fujii-2011} developed a technique for retrieving two-dimensional albedo maps of scattered light curves of face-on planetary orbits, which was extended by \citet{fujii-kawahara-2012} for general geometric configurations and was further improved by \citet{farr-etal-2018}.
\citet{fan-etal-2019} and \citet{aizawa-etal-2020} employed similar inversion techniques to study Earth-like exoplanets.
The feasibility of using techniques like this for Proxima~b was investigated by \citet{berdyugina-kuhn-2019}.

Radiation scattered by a planetary atmosphere is usually polarized \citep[e.g.,][]{kattawar-adams-1971, seager-etal-2000}, whereas the net polarization of Sun-like stars and FGK dwarfs is negligible \citep{kemp-etal-1987, cotton-etal-2017}.
Therefore, polarimetry is a promising tool for characterizing extrasolar planets \citep[e.g.,][]{stam-etal-2004, bailey-2007, stam-2008, karalidi-etal-2011, lietzow-wolf-2022}.
In particular, using scattered polarized radiation, \citet{fluri-berdyugina-2010} derived the orbital parameters and \citet{rossi-stam-2017} retrieved the cloud coverage of planets, whereas using thermally emitted polarized radiation, \citet{de-kok-etal-2011} investigated the feasibility of constraining the presence of clouds or atmospheric inhomogeneities.

This study combines the concepts mentioned in previous studies of \citet{joos-schmid-2007} and \citet{rossi-stam-2017}, but focuses on the interplay of atmospheric limb polarization with locally obscuring clouds.

The aim of this study is twofold.
We first demonstrate the utility of limb polarization of an exoplanet as a distinctive tool for extracting information about its cloud cover distribution, emphasizing the versatility and generalizability of these methods.
Second, we determine the minimum polarimetric requirements for applying these methods to a subset of planets.
This is approached as a theoretical exercise in which we propose a straightforward method, optimized for temperate rotating Jovian planets observed in edge-on systems.
We test the method on simulated phase functions.
Considering the total coverage and north-south asymmetry to be the most fundamental parameters for cloud cover distributions in a planetary atmosphere, we determine upper limits for polarimetric sensitivities necessary to distinguish between planets with different cloud cover parameters.

The study is organized as follows:
In Sect.~\ref{sec:methods-model}, we describe the numerical methods and introduce our model setup.
Subsequently, our results are presented and discussed in Sect.~\ref{sec:results} and~\ref{sec:discussion}, respectively.
Finally, in Sect.~\ref{sec:conclusions}, we summarize our study.


\section{Methods and model setup}
\label{sec:methods-model}

To calculate the scattered polarized radiation, we used the publicly available three-dimensional Monte Carlo radiative transfer code POLARIS\footnote{\url{https://portia.astrophysik.uni-kiel.de/polaris}} \citep{reissl-etal-2016} with an extension by \citet{lietzow-etal-2021} for the radiative transfer in planetary atmospheres and the local planetary environment.
The state and degree of polarization of radiation is described using the Stokes formalism \citep[e.g.,][]{bohren-huffman-1983}.
To account for a change in polarization, the Stokes vector $\boldsymbol{S} = (I, Q, U, V)^\mathrm{T}$ was multiplied by a scattering matrix $\mathbf{F}(\theta)$ at each scattering event,
\begin{equation}
    \label{eq:stokes-scattering}
    \boldsymbol{S}' \propto \mathbf{F}(\theta) \cdot \boldsymbol{S},
\end{equation}
where $\boldsymbol{S}'$ is the Stokes vector after the scattering event, and $\theta$ is the scattering angle.
Hereby, the radiation is characterized by its total flux $I$, the linear polarized fluxes $Q$ and $U$, and the circular polarized flux $V$.
In the following, we consider normalized Stokes parameters, given by
\begin{equation}
    q = \frac{Q}{I}, \quad u = \frac{U}{I}, \quad v = \frac{V}{I}.
\end{equation}
We specifically speak of net fluxes and net phase functions when we emphasize that the considered
quantities appear integrated over the effective area of an unresolved object.

For the interpretation of our results and, thus, for the description of the symmetry of the polarization phase functions, we defined a quantity as the norm of the decomposition into an even and odd part,
\begin{align}
    \label{eq:char-num}
    q_{+} &= \int_{0}^{\pi} \left\vert q(\alpha) + q(-\alpha) \right\vert\ \mathrm{d}\alpha,\\
    u_{-} &= \int_{0}^{\pi} \left\vert u(\alpha) - u(-\alpha) \right\vert\ \mathrm{d}\alpha,
\end{align}
where $q(\alpha)$ and $u(\alpha)$ are the observed phase-angle-dependent normalized net Stokes parameters.
The planetary phase angle $\alpha$ is the angle between the planet-star and planet-observer.
Thus, for $\alpha = \ang{0}$, the planet is at full phase.

Applying these norms to exoplanetary phase functions, we obtained the characteristic quantities $q_+$ and $u_-$, which serve as measures for the symmetry and antisymmetry observed in the functions $q(\alpha)$ and $u(\alpha)$, respectively.
As explained at the end of this section, phase function symmetries behave sensitively to the cloud cover distribution of a planet.
Translating these effects into scalar values allows a quantitative comparison and mapping between the cloud cover parameters and characteristic quantities.
The characteristic quantities obtained from real observational data allow constraining the corresponding cloud cover parameters, provided that the relation between parameters and quantities was established by simulations that resemble the observed system sufficiently well.

The method used for our investigation specifically targets rotating temperate Jovian exoplanets to maximize scattered flux while avoiding tidal locking.
The rotation of these gas giants allows for the assumption of latitudinal temperature gradients \citep{medvedev-etal-2013}, resulting in predominantly axially symmetric cloud patterns.
A lack of observable liquid or solid surface features additionally helps to solely attribute polarimetric effects to atmospheric origins.
Furthermore, our method requires planets from edge-on systems.
Planets like this provide a wide range of observable phase angles, while when the axial and orbital inclination are aligned, they also reveal the spatial orientation of their axially symmetric cloud cover distribution.
In terms of mass and size, our planetary model resembles a Jupiter-like planet with neglected oblateness, orbiting a Sun-like star at a distance of \SI{1}{au}.
These specific conditions are hardly represented by conventional classifications of planetary families.
Consequently, we adopted the term temperate Jovian planets, although its usage in literature is limited \citep[ e.g.,][]{alam-etal-2022, harada-etal-2023}.
However, warm Jupiters with orbital periods of $\lesssim \SI{200}{days}$ \citep[e.g.,][]{dawson-johnson-2018} are also suitable targets.

\begin{figure}
    \centering
    \includegraphics[width=\linewidth]{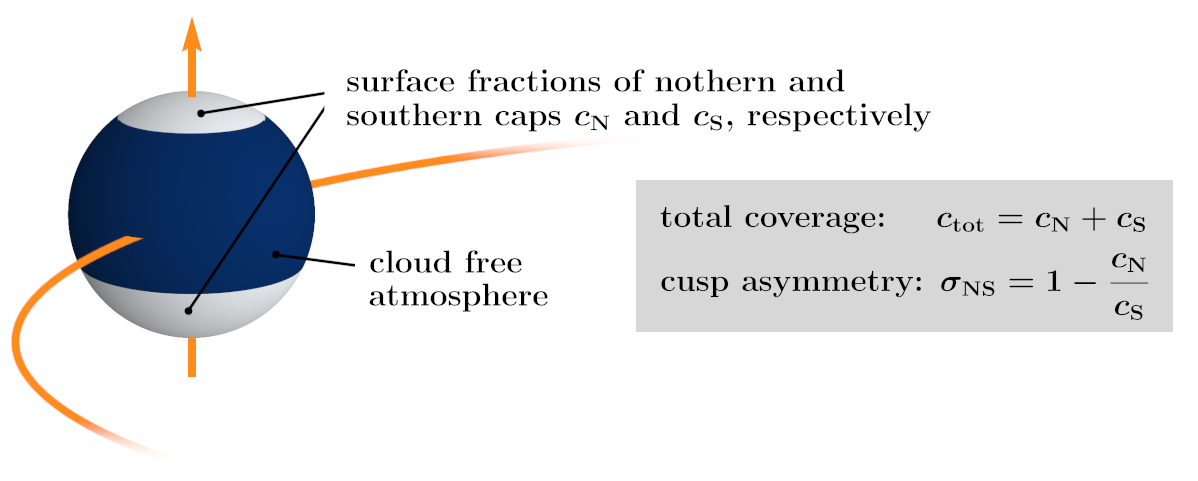}
    \caption{
        Planetary model with an exemplary cloud configuration using a total coverage of $c_\mathrm{tot} = \num{0.4}$ and a cap asymmetry of $\sigma_\mathrm{NS} = \num{0.7}$.
        The blue areas correspond to cloud-free regions, and white areas correspond to cloud-covered regions in the atmosphere.
        The observer is always located in the orbital plane.
        Oblateness, axial tilt, and orbital eccentricity are neglected.
        }
    \label{fig:planet_model}
\end{figure}

The cloud cover of Jupiter-like planets self-organizes into latitudinal systems of zones and bands.
They can either be approximated using discrete latitudinal distributions or continuous functions, such as spherical harmonics.
We assumed a discrete description of our cloud distributions as it aids in the demonstration.
However, similar to a description with spherical harmonics, we considered parameters describing a total coverage and a north-south asymmetry to be the most fundamental parameters of latitudinal cloud distributions.
The simplest realization of such a distribution assumes that clouds are confined to polar caps, which we can quantify be defining the following two parameters.
First, the total coverage $c_\mathrm{tot}$ is the fraction of the planet covered by clouds with respect to the total surface area of the planet.
Second, the north-south asymmetry $\sigma_\mathrm{NS}$ is the compliment of the ratio of the cloud coverage of the northern polar cap $c_\mathrm{N}$ compared to the cloud coverage of the southern polar cap $c_\mathrm{S}$.
Hence, a value of $\sigma_\mathrm{NS} = 0$ corresponds to symmetric polar cloud caps, while $\sigma_\mathrm{NS} = 1$ describes a planet with only southern cloud coverage.
A schematic representation of this model is shown in Fig.~\ref{fig:planet_model}.
We considered a planetary atmosphere that consisted of molecular hydrogen because that is the most abundant gas in gas giants.
The refractive index, the scattering cross section, and the scattering matrix of the gaseous particles were calculated using the formulas given by \citet{keady-kilcrease-2000}, \citet{sneep-ubachs-2005}, and \citet{hansen-travis-1974}, respectively.
The bottom pressure of the atmosphere was assumed to be \SI{100}{bar}, and thus, the total scattering optical depth amounts to about \num{2} at a wavelength of \SI{850}{nm}.
Below the atmosphere, we adopted an optically thick layer that absorbs all incident radiation.

In addition to the gaseous layers, clouds are present between pressures of \SI{1}{bar} and \SI{0.1}{bar}.
We assumed clouds that were composed of liquid water with refractive indices taken from \citet{hale-querry-1973}.
Even though the chemical environment of clouds is expected to be quite complex in gas giants, we simplified our model by the choice of a single cloud species.
The cloud particles were described using the size distribution by \citet{hansen-1971},
\begin{equation}
    n(s) \propto s^{(1 - 3 v_\mathrm{eff}) / v_\mathrm{eff}}\ \exp[-s / (s_\mathrm{eff} v_\mathrm{eff})],
\end{equation}
where $s_\mathrm{eff}$ and $v_\mathrm{eff}$ are the mean effective radius and effective variance, respectively.
We considered an effective radius of \SI{1}{\um} and an effective variance of \num{0.1}, similar to previous studies \citep[e.g.,][]{stam-etal-2004, karalidi-etal-2013}.
The cross sections and scattering matrix were calculated assuming spherical particles using the code \texttt{miex} \citep{wolf-voshchinnikov-2004}, which is based on the Mie scattering theory \citep{mie-1908}.
For the optical depth of the cloud layer, we adopted a value of \num{30} at a wavelength of \SI{850}{nm} to reflect the large optical depth of the clouds on Jupiter that was assumed in previous studies \citep{sromovsky-fry-2002, mclean-etal-2017}.

\begin{figure*}
    \centering
    \includegraphics[width=\linewidth]{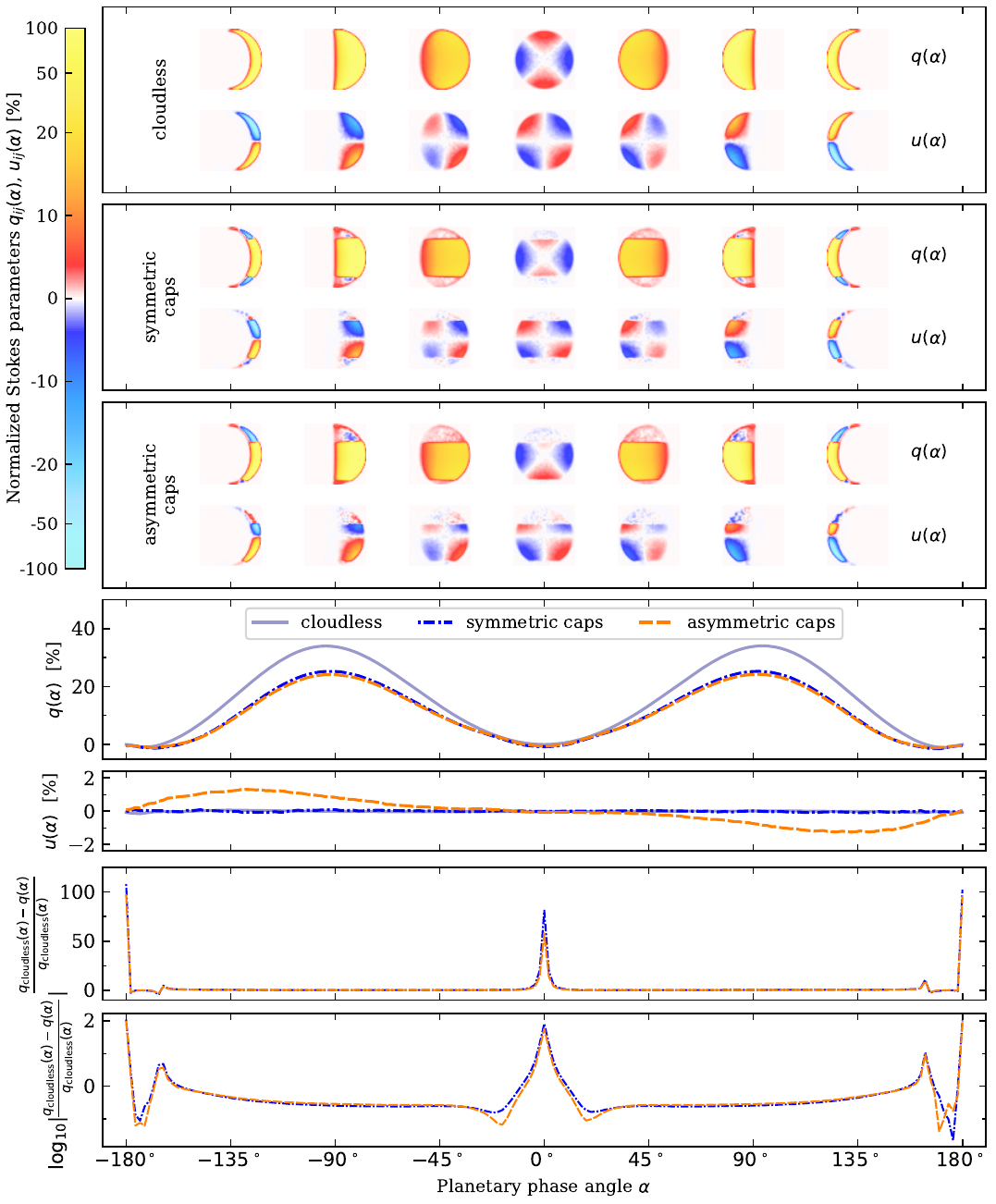}
    \caption{
        Polarization of planetary phase functions for the cases of two cloud cap configurations and for a cloud-free atmosphere to illustrate the effect of locally obscured limb polarization on the net phase function at wavelengths of $\lambda = \SI{550}{\nm}$.
        The upper three panels show the spatially resolved normalized Stokes parameters $q(\alpha)$ and $u(\alpha)$ for selected phase angles $\alpha$ of spatially resolved planetary disks.
        Below, the resulting net normalized Stokes parameters $q(\alpha)$ and $u(\alpha)$ are shown.
        The bottom two panels show the relative differences of $q(\alpha)$ and the reference case of a cloudless atmosphere at a linear and logarithmic scale.
        For all figures, the phase angles range from \ang{-180} to \ang{180}.
        A full planetary orbit is therefore considered.
    }
    \label{fig:phase_function_examples}
\end{figure*}

In planetary atmospheres, a phenomenon referred to as limb polarization arises due to multiple Rayleigh scattering by small particles, leading to a high degree of linearly polarized radiation received from the limb of the planet \citep[see, e.g.,][]{schmid-etal-2006, joos-schmid-2007}.
When optically thick cloud layers locally obstruct this phenomenon, their positional information is encoded in the resulting polarization of the net flux of the planet.
Figure~\ref{fig:phase_function_examples} illustrates this principle for two different cloud configurations and for a cloud-free atmosphere.

The amplitudes of the symmetric net contributions to normalized Stokes $q(\alpha)$ display an anticorrelation with the total cloud coverage of the planet, as was similarly found by \citet{rossi-stam-2017} for planetary models with polar caps.
However, the differences in the north-south distribution of the polar caps most notably produce antisymmetric net contributions to the polarization $u(\alpha)$.
The spatially resolved polarization maps of the radiation scattered by the planet reveal that asymmetric blocking of the limb polarization by asymmetric polar cloud caps causes this behavior.
For our model, we found this effect to be most pronounced at a wavelength of about \SI{850}{nm}.
At shorter wavelengths, the incoming radiation is predominantly scattered at higher altitudes, which decreases the influence of deeper-lying cloud layers.

\begin{figure}
    \centering
    \includegraphics[width=0.9\linewidth]{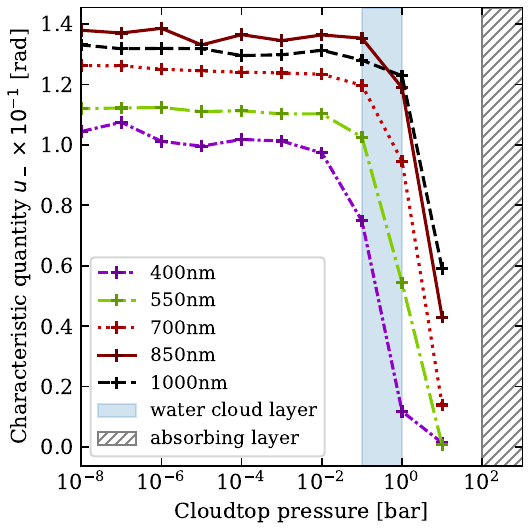}
    \caption{
        Characteristic quantities $u_{-}$ for various wavelengths and cloud layer heights.
        The planetary model has a total cloud coverage of $c_\mathrm{tot} = \num{0.5}$ and a cap asymmetry of $\sigma_\mathrm{NS} = \num{1}$.
        The blue area marks the cloud layer pressure assumed throughout this study.
        The hatched black area represents an absorbing layer, which is the lower boundary of our model space.
    }
    \label{fig:wavelengths_cloudtoppressures}
\end{figure}

We tested the above dependence on wavelength and height of the cloud layer by simulating phase functions between \SI{400}{nm} and \SI{1000}{nm} for planet models with cloud layers located at different cloud-top pressures.
The models share pronounced polar cap asymmetries ($c_\mathrm{total} = \num{0.5}$, $\sigma_\mathrm{NS} = \num{1}$), and therefore, they produced specific values of the characteristic quantities $u_-$.
We took the resulting values of $u_-$ as a qualitative measure of the efficiency with which the limb polarization reveals the cloud cover distributions.
The results are shown in Fig.~\ref{fig:wavelengths_cloudtoppressures} for different wavelengths and heights of the cloud layer.

The lower two plots in Fig.~\ref{fig:phase_function_examples} show the relative differences of the phase function $q(\alpha)$ of the considered cloud configurations with the cloudless case as reference.
This indicates that a further differentiation between the considered phase functions is indeed possible.
Notable differences between the polar cap models appear to be mostly confined to phase angles near \ang{0} and \ang{\pm 180}.
These phase angles are especially sensitive to more complex differences in latitudinal cloud cover distributions.
However, accounting for these distribution properties would require additional cloud cover parameters, for instance, a poles-to-equator ratio.
When the distributions were to be characterized via spherical harmonics, this would correspond to the use of expansion coefficients of progressively higher order.
It is worth noting that $q(\alpha)$ is responsive to even coefficients, starting with coefficient zero resembling the total coverage, while $u(\alpha)$ is responsive to odd coefficients, starting with the first coefficient resembling the north-south asymmetry.
Nevertheless, these influences decrease with higher coefficient orders and are progressively harder to separate in the affected phase functions.
Therefore, and with the goal of finding a meaningful upper limit at which polarimetric precision suffices to extract the most fundamental information about the cloud cover, we limited ourselves to the total coverage and the north-south asymmetry represented by the characteristic quantities $q_+$ and $u_-$.

Eventually, the polarimetric precision $\Delta p$ of an instrument is a fundamental technical limitation that affects all quantities derived subsequently.
Thus, according to an assumed precision, the proposed characteristic quantities have uncertainty intervals $q_{+} \pm \Delta q_{+}$ and $u_{-} \pm \Delta u_{-}$ that define whether we can distinguish between different cloud configurations.
A total of \num{57} cloud configurations were examined, ranging from cloud-free atmospheres to full cloud coverages and from symmetric to highly asymmetric polar caps.
We modeled the cloud configurations for any parameter combination of $c_\mathrm{tot}$ and $\sigma_\mathrm{NS}$ between \num{0} and \num{1}.
The respective step widths were \num{0.1} for $\sigma_\mathrm{NS}$ and \num{0.2} for $c_\mathrm{tot}$ within $0.1 \leq c_\mathrm{tot} \leq 0.9$.

For each cloud configuration, we simulated the three-dimensional radiative transfer in the planetary atmosphere and calculated the state and degree of the normalized polarization of the phase-angle-dependent scattered radiation.
Subsequently, using the normalized net fluxes of the Stokes parameters $q(\alpha)$ and $u(\alpha)$, we computed pairs of characteristic quantities according to Eq.~\eqref{eq:char-num}.


\section{Results}
\label{sec:results}

\begin{figure*}
    \centering
    \includegraphics[width=\linewidth]{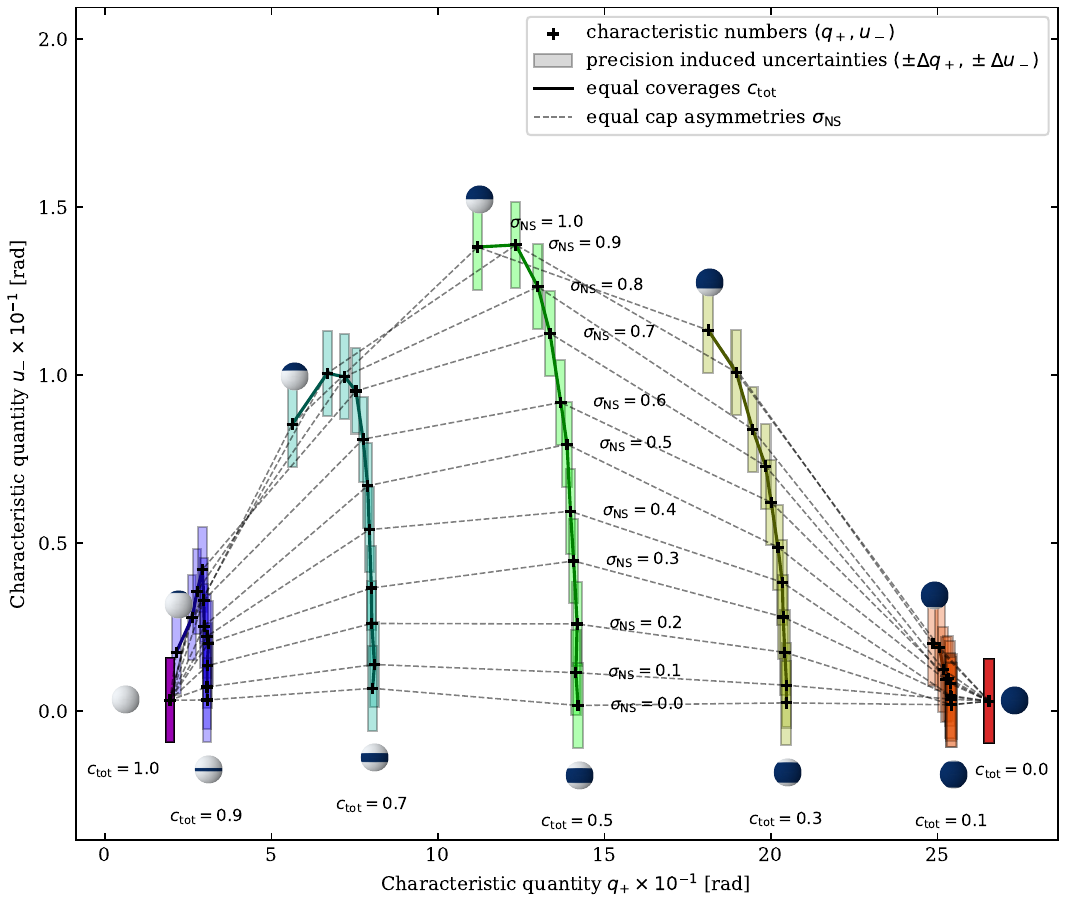}
    \caption{
        Characteristic quantities $q_{+}$ and $u_{-}$ for all 57 considered cloud configurations under the assumption that the planetary signal can be separated from the stellar radiation (stellar radiation excluded) at an optimal wavelength of $\lambda = \SI{850}{\nm}$.
        The tuples of the characteristic quantities with equal total cloud coverages $c_\mathrm{tot}$ are connected by solid lines and are represented by the same color.
        The tuples of the characteristic quantities with equal cap asymmetries $\sigma_\mathrm{NS}$ are connected by dashed lines.
        The error boxes indicate the uncertainties as expected for a polarimetric precision of $\Delta p = \SI{0.1}{\percent}$.
        }
    \label{fig:char_num_star_excluded}
\end{figure*}

Figure~\ref{fig:char_num_star_excluded} shows the characteristic quantities $q_+$ and $u_-$ for the case in which the stellar radiation is not included in the net flux.
Each point in the diagram represents a tuple of characteristic quantities derived from a pair of phase functions in normalized Stokes parameters $q(\alpha)$ and $u(\alpha)$ and considers a specific cloud configuration.
Distinct cloud configurations sharing the same cloud coverage $c_\mathrm{tot}$ are connected by solid lines.
Similarly, characteristic quantities of cloud configurations with identical cap asymmetries $\sigma_\mathrm{NS}$ are connected by dashed lines.
The error boxes around each individual tuple of the characteristic quantities denote the uncertainty of their measurements due to an assumed polarimetric precision $\Delta p$.
To distinguish the tuples of the characteristic quantities and to distinguish the considered cloud configurations, $\Delta p \leq \SI{0.1}{\percent}$ was required when the stellar radiation was not included in the net flux.

\begin{figure*}
    \centering
    \includegraphics[width=\linewidth]{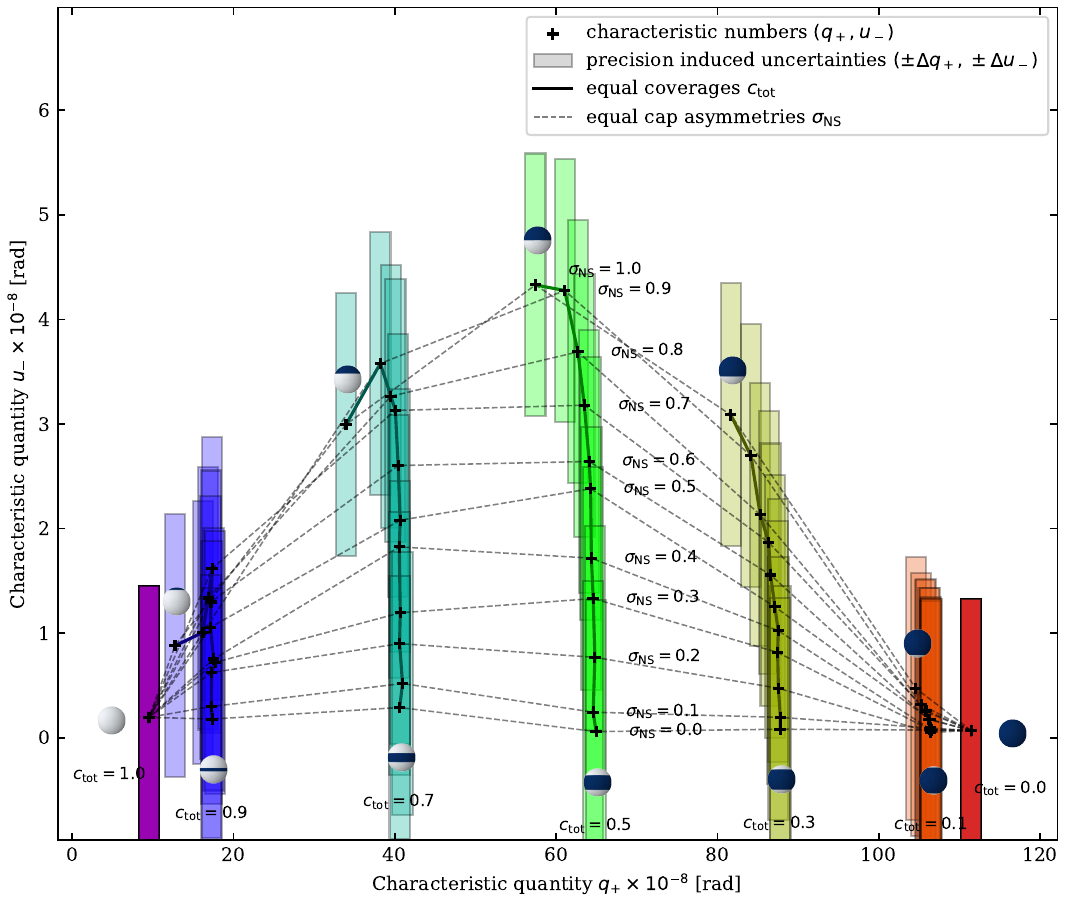}
    \caption{
        Same as Fig.~\ref{fig:char_num_star_excluded}, but with stellar radiation.
        The error boxes indicate the uncertainties expected for a polarimetric precision of $\Delta p = \SI{e-3}{ppm}$.
        }
    \label{fig:char_num_star_included}
\end{figure*}

In contrast, Fig.~\ref{fig:char_num_star_included} shows the characteristic quantities for the case in which the stellar radiation is included in the net flux.
It would require a polarimetric precision in the range of $\SI{e-2}{ppm}$ to $\SI{e-3}{ppm}$ for such a system to reveal information about the cloud configuration of the planetary atmosphere.

The distribution of the characteristic quantities matches our expectations and can be explained as follows.
The dependence of $q_{+}$ on the total coverage $c_\mathrm{tot}$ is most notable.
It forms prominent hook-shaped columns scattered along the abscissa.
Higher values of $q_{+}$ are strongly correlated with decreasing cloud coverages.
This is the result of less extensive cloud coverages, enabling underlying gas layers to scatter the incoming stellar radiation directly toward the observer via single Rayleigh-scattering events.
The effective fractional cloudless area of the planet, which enables these scattering events, is the most prominent contributor to the symmetric shape of the phase function in Stokes parameter $q(\alpha)$ (see Sect.~\ref{sec:methods-model} and \citealt{rossi-stam-2017}).
This is shown in Fig.~\ref{fig:phase_function_examples}.
It results in pronounced maxima at $\alpha = \pm \ang{90}$.

In contrast, the characteristic quantities $u_{-}$ are most sensitive to the polar cap asymmetry $\sigma_\mathrm{NS}$, especially at medium coverages around $c_\mathrm{tot} = \num{0.5}$.
As illustrated in Fig.~\ref{fig:phase_function_examples}, symmetric cap coverages cause the normalized Stokes parameter $u(\alpha)$ to remain zero for all planetary phase angles $\alpha$, while cap asymmetries introduce antisymmetric behaviors of $u(\alpha)$.
This is to be expected because in the case of symmetric cloud caps, the contributions of polarized fluxes from the northern and southern hemispheres originating from the limb of the planet cancel each other out.
However, when one hemisphere is disproportionately covered by clouds, it enables the flux of the other hemisphere to contribute to a net polarized phase function, which is antisymmetric for normalized Stokes parameter $u(\alpha)$.
As a direct consequence, the characteristic quantities $u_{-}$ remain zero for all cloud configurations with symmetric polar caps ($\sigma_\mathrm{NS} = \num{0}$), while for more asymmetric caps ($\sigma_\mathrm{NS} > \num{0}$), the characteristic quantities $u_{-}$ deviate from zero.
At both higher and lower cloud coverages, a decreasing fraction of either a clear atmosphere or cloud layers results in lower polarization asymmetries across the visible planetary disk.
Thus, despite the cap asymmetries, the characteristic quantities $u_{-}$ decrease for both higher and lower $\sigma_\mathrm{NS}$, culminating in $u_{-} = \num{0}$ for the extreme cases of a cloudless planet ($c_\mathrm{tot} = \num{0}$) as well as a planet that is completely covered by clouds ($c_\mathrm{tot} = \num{1}$).

Finally, the formation of hook-shaped distributions with increasing cloud coverages can be explained by the fact that as the cloud coverage exceeds $c_\mathrm{tot} = \num{0.5}$, a cloud configuration is characterized by a progressively narrower band of a cloud-free atmosphere.
In this narrow region, limb polarization can occur, providing an asymmetric contribution to the normalized Stokes parameter $u(\alpha)$.
With increasing cap asymmetry $\sigma_\mathrm{NS}$, the cloudless band moves closer to the poles of the planet.
However, since the most effective contribution of limb polarization to an antisymmetric net function of $u(\alpha)$ occurs at intermediate latitudes, the characteristic quantities $u_{-}$ will fluctuate as the cloud-free atmospheric region crosses these mid-latitudes toward the poles of the planet.
Consequently, the highest cap asymmetries $\sigma_\mathrm{NS}$ do not necessarily cause the highest antisymmetric distribution of the polarized Stokes parameter $u(\alpha)$ or the highest values of $u_{-}$ when the total cloud coverage amounts to $c_\mathrm{tot} \geq \num{0.5}$.


\section{Discussion}
\label{sec:discussion}

We showed that the feasibility of polarimetric approaches for retrieving latitudinal cloud cover information of exoplanetary atmospheres strongly depends on the direct stellar flux.
When the stellar radiation is not considered in the net flux, a polarimetric precision of \SI{1}{\percent} to about \SI{0.1}{\percent} is required to retrieve the most fundamental information about the latitudinal cloud cover distributions.
In contrast, a precision of $\SI{e-2}{ppm}$ to about $\SI{e-3}{ppm}$ is required when the direct stellar emission is included in the polarimetric measurements, which is higher by about a factor of about $\num{e2}$ to $\num{e3}$ than is currently achievable with polarimeters such as the High-Precision Polarimetric Instrument-2 \cite[HIPPI-2;][]{bailey-etal-2020}.
Future surveys, for instance, the Large UV/Optical/Infrared Surveyor \citep[LUVOIR;][]{luvoir-team-2019} mission, are promising steps toward star-independent observations of polarized phase functions.
However, POLLUX \citep{muslimov-etal-2018}, a high-resolution spectropolarimeter for the future LUVOIR mission, examines the UV to visible wavelength range, where the gaseous layers above the clouds are too optically thick for our method to work efficiently.
In the context of our model, wavelengths in the range of $\SI{500}{nm}$ to about $\SI{1000}{nm}$ proved to be much more promising.

The above prospects have to be considered in context because planetary and atmospheric systems are far more complex than the simplified model we used for the illustration of this method in this study.
Within the framework of our model, we find a broadly expressed bijective correspondence between the two characteristic quantities and cloud model parameters.
While this allows for a convenient retrieval of cloud cover information, the introduction of additional model parameters would require supplementary characteristic quantities to avoid ambiguities.
Orbital inclination and axial orientation are important examples that significantly impact symmetric and asymmetric phase function behaviors.
Additional polarimetric influences due to moons \citep{berzosa-molina-etal-2018}, circumplanetary rings \citep{lietzow-wolf-2023}, or further planets in a multiplanetary systems have to be either considered or excluded from the data if possible.
In particular, neighboring planets could be filtered out through frequency domain techniques.
However, prominent rings and large moons pose a much greater challenge as radiative interaction with their host planet can be quite complex \citep[see, e.g.,][]{berzosa-molina-etal-2018, lietzow-wolf-2023}.

Nonetheless, numerous unknowns can be constrained, determined, or mitigated.
For instance, when it is limited to edge-on orbits, transit spectroscopy can reveal the chemical composition of the exoplanetary atmosphere and its cloud layers.
Furthermore, the vertical structure of the atmosphere can be probed by expanding our polarimetric approach to multiwavelength observations.
As in the case of Rayleigh scattering, the scattering cross section decreases for an increasing wavelength of the radiation.
As discussed in Fig.~\ref{fig:wavelengths_cloudtoppressures}, this causes observations conducted at shorter wavelengths to primarily expose cloud features at higher atmospheric altitudes, which can then be compared to longer-wavelength observations that reach deeper layers.

Another fundamental constraint to our model is the assumption of polar caps.
Although the axially symmetric distribution of the cloud cover is consistent with the behavior of rotating gas giants, polar caps represent a more extreme subset of cases.
The choice of polar caps is motivated by the simplicity of their parameterization regarding coverage and asymmetry as well as their optimal impact on the phase-angle-dependent polarization of light scattered by the planet, aiding the purpose of showing the feasibility of this concept.
Another conceivable approach would be to use zonal coefficients of spherical harmonics, which act as a finite but expandable set of parameters to model cloud cover distributions at a desired degree of accuracy.

When sufficiently highly resolved planetary phase functions are available, polarimetric approaches will offer a variety of opportunities to help us to study and map exoplanets by linking many existing methods or those under development.


\section{Conclusions}
\label{sec:conclusions}

We demonstrated the potential of polarimetric techniques for retrieving surface distributions of exoplanetary cloud covers by analyzing the state and degree of polarization of the scattered radiation.

Our work complements existing considerations with a specific emphasis on the interplay between the cloud coverage of a planet and its limb polarization.
Using this effect, we determined an upper limit for the polarimetric precision required to constrain the cloud coverage.
For this purpose, we proposed a straightforward method that has proven most effective for temperate Jovian planets in edge-on systems and uses characteristic quantities derived from the normalized net linear polarization of the planetary phase functions.
These quantities are highly sensitive to their cloud cover distribution, in particular, to the cloud coverage and cap asymmetry.
Following this method, our results indicate that cloud cover information can be retrieved at a polarimetric precision of \SI{1}{\percent} when the planetary signal is observed independently of its host star.
Considering observations that include both reflected planetary flux and stellar flux, the required polarimetric precision drops to 
$\SI{e-3}{ppm}$, which lies about three magnitudes below current technical capabilities.

In summary, this work contributes to the emerging field of exocartography and highlights the complementary role of polarimetry alongside traditional photometric and spectroscopic techniques.
We hope that our method will inspire future ideas for analyzing the polarization of exoplanets, taking advantage of the promising opportunities that increasingly accurate polarimeters will offer.


\begin{acknowledgements}

The authors thank the anonymous referee for useful comments and suggestions.
This research made use of
\href{https://www.astropy.org}{Astropy} \citep{astropy-etal-2013, astropy-etal-2018, astropy-etal-2022},
\href{https://matplotlib.org}{Matplotlib} \citep{Hunter-2007},
\href{https://numpy.org}{Numpy} \citep{Harris-etal-2020},
and \href{https://ui.adsabs.harvard.edu}{NASA's Astrophysics Data System}.

\end{acknowledgements}


\bibliographystyle{aa-link}
\bibliography{bibliography.bib}

\end{document}